\documentclass[conference]{IEEEtran}
\IEEEoverridecommandlockouts
\usepackage{amsmath,amsfonts}
\usepackage{algorithmic}
\usepackage{array}
\usepackage{textcomp}
\usepackage{stfloats}
\usepackage{url}
\usepackage{verbatim}
\usepackage{graphicx}
\usepackage{cite}
\usepackage{amssymb}
\usepackage{color,colortbl}
\usepackage{pgfplots}
\usetikzlibrary{calc}
\usepackage{multirow}
\usepackage{circuitikz}
\usepackage{subcaption}
\usepackage{siunitx}
\usetikzlibrary{shapes}
\usepackage{tikz}
\usepackage{environ}
\makeatletter
\newsavebox{\measure@tikzpicture}
\NewEnviron{scaletikzpicturetowidth}[1]{%
  \def\tikz@width{#1}%
  \begin{lrbox}{\measure@tikzpicture}%
  \BODY
  \end{lrbox}%
  \pgfmathparse{#1/\wd\measure@tikzpicture}%
  \BODY
}

\def\BibTeX{{\rm B\kern-.05em{\sc i\kern-.025em b}\kern-.08em
    T\kern-.1667em\lower.7ex\hbox{E}\kern-.125emX}}

\usepackage[normalem]{ulem}\usepackage{xspace}

\begin{document}

\title{Nonlinear Stacked Intelligent Surfaces for Wireless Systems}
\author{Omran Abbas, Abdullah Zayat, Lo\"{i}c~Markley, and Anas Chaaban
\thanks{The authors are with the School of Engineering, University of British Columbia, Kelowna, Canada (e-mail: omran.abbas@ubc.ca, abduzaya@ubc.ca loic.markley@ubc.ca, anas.chaaban@ubc.ca).}}
\maketitle

\begin{abstract}

Stacked intelligent surfaces (SIS) are a promising technology for next-generation wireless systems, offering an opportunity to enhance communication performance with low power consumption. Typically, an SIS is modelled as a surface that imparts phase shifts on impinging electromagnetic signals to achieve desired communication objectives. However, this mode of operation results in a linear SIS, which limits its applicability to linear operations. To unlock further SIS potential, we propose a nonlinear SIS that can mimic the behaviour of nonlinear neural networks. We discuss the feasibility and potential of this idea and propose a nonlinear SIS unit cell with a step-like response. To evaluate the system-level performance of nonlinear SIS, we present a case study where SIS structures are optimized to minimize the symbol error rate (SER) in an MIMO system with SIS deployed at both the transmitter and receiver sides using only statistical channel information. We demonstrate that a nonlinear SIS can improve communication reliability compared to a linear SIS by forming complex signal patterns across the SIS surface, which provide higher diversity against noise disturbances, while still allowing the receiver to discern these patterns. Finally, we outline several potential applications of nonlinear SIS in wireless communication scenarios.

\end{abstract}



\section{Introduction}

Future wireless communication systems are envisioned to support emerging use cases such as holographic telepresence and multisensory experiences, in addition to the growing demands of existing 5G applications \cite{6GSurvey}. Realizing these visions necessitates the adoption of advanced technologies, such as ultra-massive multiple-input multiple-output (mMIMO) and integrated sensing and communication (ISAC), to address challenges stemming from severe link attenuation and spectral congestion. However, these solutions often incur high energy consumption, reaching several hundred watts \cite{challeng6G}, alongside increased transceiver complexity and radio frequency (RF) hardware-induced nonlinearities due to the adoption of ultra-wideband signalling \cite{6GSurvey}. In response to these limitations, stacked intelligent surfaces (SIS) have emerged as a promising evolution of reconfigurable intelligent surfaces (RIS), offering new avenues for enhancing wireless system performance \cite{SISComSens}.

An SIS comprises multiple layers of passive or active elements arranged in front of a transceiver and operating in the transmissive mode to change the phase and/or amplitude of incident electromagnetic (EM) waves. This architecture enables direct wave-domain manipulation for functions such as MIMO precoding, interference suppression, and beamforming \cite{An_SIM_transceiver}. By doing so, an SIS can effectively replace conventional RF circuitry with a cost-efficient and largely passive structure, enabling ultra-fast signal processing for both sensing and communication tasks, while significantly reducing system complexity, energy consumption, and hardware cost. Various studies have demonstrated the utility of SIS in wireless systems for communication and sensing applications \cite{gradientdecentSIMComm}. These functionalities are typically achieved by optimizing the phase shift imparted by each unit cell across the SIS.

\subsection{SIS as Neural Network}

The SIS can be interpreted as a neural network (NN), highlighting its potential for enabling wave-domain artificial intelligence (AI)-based physical layer functionalities, akin to earlier implementations of digital AI in communication systems \cite{SIM_NN}. The relevance of AI in future physical layer design stems from the increasing departure from classical system assumptions, driven by emerging technologies that introduce significant nonlinearities and complexity, rendering optimal solutions analytically intractable and existing solutions suboptimal.
Despite its potential, SIS's ‘intelligent’ nature remains largely untapped. Most existing works deploy SIS as a spatial linear wave-domain processor, primarily for beamforming. Although some studies have modelled SIS as an NN \cite{SIM_NN}, the resulting architectures are constrained to linear operations. This imposes a fundamental limitation: it is well-known that linear NNs, regardless of depth, cannot solve even elementary nonlinear tasks, such as the XOR problem. Moreover, from both mathematical and expressive standpoints, stacking linear layers offers no advantage over a single linear transformation.

\subsection{State of the Art}

Typically, research on the application of intelligent surfaces to wireless communications focuses on passive structures, which are favoured for their inherently low power consumption \cite{SISComSens}. Such designs typically optimize only the phase shifts introduced by the surface elements, though some studies have also considered element losses and their impact on system performance \cite{lossyRIS}.
More recent efforts, however, have explored surfaces incorporating active elements, i.e., components capable of imparting both phase shifts and power gain, to enhance system performance.
This amplifying functionality introduces a form of nonlinearity (albeit at the component level), as it often involves the integration of nonlinear components such as transistors. For instance, the authors in \cite{RISActive} demonstrated an active RIS with amplifying unit cells that mitigates the double path-loss effect, significantly improving achievable rates compared to its passive counterpart. Additionally, the study in \cite{ActiveSIM} investigated an active SIS in a downlink multi-user setting, where active and passive layers were jointly utilized. These architectures achieved notable gains in spectral efficiency over conventional SIS designs by simultaneously optimizing the phase shifts and the amplification gains under power constraints.

\subsection{Our Contribution}

In this work, we explore SIS architectures that extend beyond conventional SIS with linear processing by introducing nonlinear functionality into the design. A nonlinear SIS (NL-SIS) can be interpreted as a nonlinear NN, thereby unlocking the potential for wave-domain AI to perform physical layer functions with embedded intelligence. This paradigm has the potential to substantially enhance the performance of wireless systems while maintaining reduced transceiver complexity and energy consumption. Unlike existing approaches that deploy active elements to achieve amplification and compare them against passive implementations, we propose a design that employs passive elements (characterized by zero amplitude gain and low power consumption) augmented with conventional phase control and a nonlinear amplitude response. This approach provides two primary advantages: (i) it introduces a nonlinear amplitude response as an additional degree of freedom without compromising energy efficiency, and (ii) it enables deep AI contrary to stacking linear layers, thereby facilitating more complex operations with enhanced noise robustness.

This paper lays the groundwork for the NL-SIS concept, demonstrating its practical feasibility and highlighting its advantages in wireless communication systems. In particular, in this paper,
\begin{itemize}
    \item we demonstrate the feasibility of NL-SIS by examining multiple nonlinear surface designs and proposing a conceptual unit cell architecture for an NL-SIS;

    \item we analyze the system-level potentials of NL-SIS-assisted systems and use a relevant optimization strategy tailored to their nonlinear nature, highlighting the distinctions from conventional SIS optimization;

    \item we present a case study showcasing the symbol error rate (SER) enhancement and noise resilience enabled by NL-SIS; and
    \item we discuss prospective applications and challenges, outlining a vision for future research and development of NL-SIS.
\end{itemize}

In the following section, we delve into the NL-SIS architectural framework, operational principles, and potential physical realization.

\section{Nonlinear Surface}\label{Sec.Nonlinear}

Engineered surfaces have been a popular approach for the efficient control of EM waves, enabling various applications such as absorbers, polarizers, and holographic plates \cite{ActiveSIM,active_non_lin_2}. These have included diffraction gratings, reflectarrays, and even metasurfaces. Fundamentally, these surfaces consist of periodic arrays of engineered unit cells designed to scatter EM fields with precision and purpose.

In this section, we outline the concept of nonlinear surfaces and their capabilities through several applications. We then propose an architectural design for nonlinear surfaces that extends the functionality of conventional SIS, currently limited to linear NN behaviour, into an NL-SIS framework.

\subsection{Review of Linear vs Nonlinear NN}\label{nl_SIM_theory}

Nonlinearity is a fundamental concept in NNs, serving as the basis for enabling complex and expressive functionalities. In NNs, nonlinearity is typically introduced through activation functions at the neuron level (sigmoid, ReLU, etc.), allowing the network to learn nontrivial mappings between inputs and outputs. Without nonlinear activation, an NN, regardless of its depth, collapses into an equivalent linear transformation, thus severely limiting its representational capacity.

The same principle can be translated into the SIS, where each unit cell plays the role of a neuron. In a conventional SIS, each unit performs a linear transformation (typically a phase shift) on the incident EM wave, which restricts the overall surface behaviour to a linear mapping. To unlock the full potential of SIS as a physical-layer NN, nonlinearity must be introduced at the unit cell or layer level. This can be achieved by designing elements whose EM response is amplitude-dependent, thus mimicking nonlinear activation functions in NNs. The nonlinearity allows the SIS to perform complex nonlinear transformations between input and output wavefronts, which is otherwise unattainable in purely linear structures, all while preserving low operational and manufacturing costs.

Toward that goal, two fundamental questions arise: (i) what is a feasible architecture for a nonlinear unit cell, and (ii) is such a design beneficial in practical systems? This section addresses the first question through examples; the second will be explored in the subsequent section.

\subsection{Review on Linear vs Nonlinear Components and Unit Cells}

To clarify the different forms of nonlinearity in surfaces, it is essential to differentiate between linear and nonlinear components and linear and nonlinear unit cells:

\begin{itemize}

\item[(i)] A linear component exhibits a proportional relationship between input and output variables such as voltage and current, e.g., resistors, capacitors, and inductors.
\item[(ii)] A nonlinear component exhibits a non-proportional response between the input and the output, e.g., diodes, varactors, and transistors. Note that many nonlinear components can be linearized under small signal conditions; to ensure nonlinearity, a nonlinear component must be operated beyond its linear region.

\item[(iii)]A linear unit cell refers to a surface element that exhibits a linear response to incident EM waves. This unit cell may contain linear and nonlinear components, provided the nonlinear components are operating within their linear region. For instance, a linear unit cell can contain varactors to control the phase shift in its linear input-output wave relationship.

\item[(iv)] A nonlinear unit cell exhibits a nonlinear response to an incident EM wave by embedding components that are nonlinear \textit{and} operating in their nonlinear region. For instance, a unit cell that switches between absorbing and transmitting modes depending on signal intensity produces a nonlinear input-output relationship. Such surfaces can be categorized based on their activation and control, as explained in the next subsection. 
\end{itemize}

The integration of nonlinear components into reconfigurable surfaces has been explored in several studies \cite{RISActive,ActiveSIM}. For example, the authors in \cite{RISActive} demonstrate that integrating an amplifying RIS into a multiple-input single-output system enhances capacity and outage performance in low-amplitude scenarios with imperfect channel state information (CSI). In the context of SIS \cite{ActiveSIM}, combining active and passive layers with a discretized phase enhanced the achievable rates, yielding a beamforming gain comparable to that of an SIS with continuous phase control. These works mark an important milestone in reconfigurable surfaces research; however, nonlinearity is implemented at the component level, but is not fully exploited in the operation and optimization framework at the unit cell level. Specifically, these surfaces provide power gain to reflected or transmitted waves without incorporating a nonlinear input-output relation.

\subsection{Feasibility of Nonlinear Surfaces}

An SIS is considered nonlinear when its input-output relation has a nonlinear behaviour. Accordingly, an SIS with unit cells that are constrained to scale an incident EM wave (i.e. applying a phase shift or introducing amplification) is not considered an NL-SIS.

One way to realize nonlinearity in an SIS is by using a control mechanism to configure the surface response so it has the behaviour of a nonlinear unit cell. For instance, a nonlinear configuration is presented in \cite{active_non_lin_2} that incorporates a tunable transmissive element, EM detectors which convert the EM wave power into current, and a digital control module which processes the detector currents and generates a controlling digital signal. This digital control module adjusts the element functionality to either transmit or absorb the EM waves based on the detected power level.

Alternatively, an NL surface can be realized by incorporating a passive element or material whose response varies nonlinearly with the frequency, power, or polarization of the impinging EM waves.
The study in \cite{self_nonlinear} investigated a nonlinear surface employing PIN diodes, where the surface response is configured based on the intensity of a linearly polarized incident wave. The proposed design features unit cells composed of two anti-phase elements, each integrated with a PIN diode. At low incident wave intensities, the PIN behaves as an open circuit, resulting in polarization conversion with full reflection. As the intensity increases, a sufficient voltage is induced across the PIN terminals and it transitions to an inductive state, inducing cross-polarized reflection and effectively functioning as a digital spatial wave manipulator \cite{self_nonlinear}.

\subsection{Potential Architecture}

From a modelling and optimization perspective, embedding a nonlinearity in an SIS akin to the nonlinearity in a nonlinear NN can be realized using a unit cell that performs an amplitude-dependent operation on the incident EM wave. In this context, a unit cell exhibiting an activation function similar to a step function presents a promising design option. Considering a signal of the form $x(t)=a(t)e^{j\phi(t)}$ where $a(t)\geq 0$, an amplitude-step function of the form $f(x(t))={\rm Step}(a(t)+T) x(t)e^{j\theta}$, where $T\leq 0$ is a tunable threshold, $\theta$ is a tunable phase shift, and ${\rm Step}(\cdot)$ is the unit step function, achieves the desired nonlinear amplitude-dependent functionality while preserving the ability to independently change the phase of $x(t)$. Notably, this is a passive operation, i.e., it does not incur additional power. As we shall see, it can be realized with low complexity.

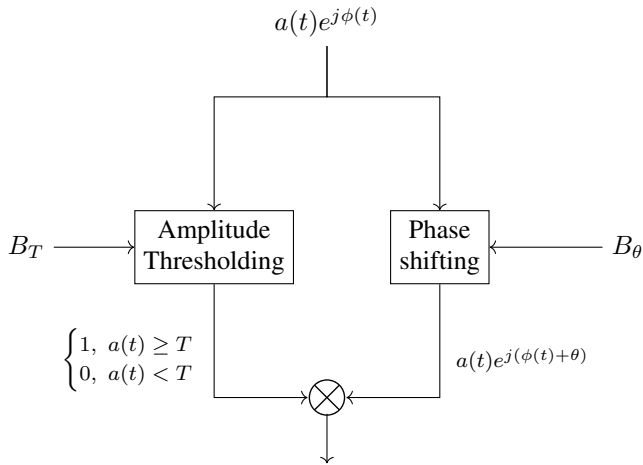
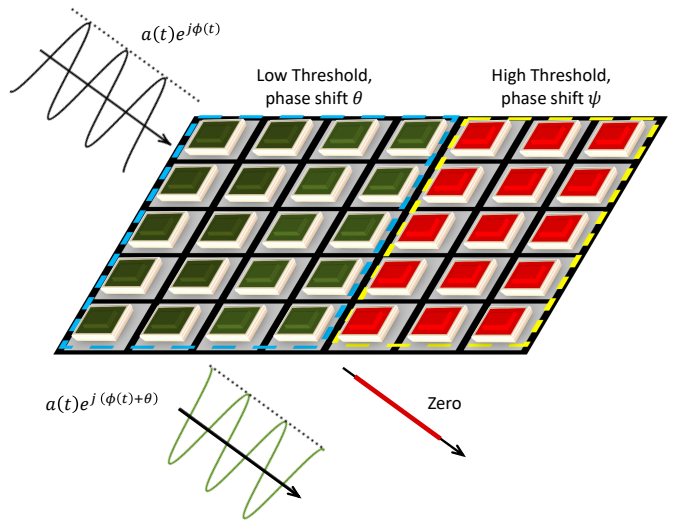
\begin{figure*}[htbp]
    \centering
    \begin{subfigure}{0.49\textwidth}
       \tikzset{every picture/.style={scale=1}, every node/.style={scale=1}}
        \begin{tikzpicture}
\node (in) at (0,0) {$a(t)e^{j\phi(t)}$};
\node (c1) at (0,-1) {};
\node[draw,align=center] (at) at (-1.5,-3) {Amplitude \\ Thresholding};
\node[draw,align=center] (ps) at (1.5,-3) {Phase \\ shifting};
\node (t) at (0,-5) {\huge $\otimes$};
\node (o) at (0,-6) {};
\node (bt) at (-4,-3) {$B_T$};
\node (bp) at (4,-3) {$B_{\theta}$};

\draw [->] (in) to (c1.center) to ($(c1.center)-(1.5,0)$) to (at);
\draw [->] (in) to (c1.center) to ($(c1.center)+(1.5,0)$) to (ps);
\draw [->] (at) to ($(t.center)-(1.5,0)$) to ($(t.west)+(.15,0)$);
\draw [->] (ps) to ($(t.center)+(1.5,0)$) to ($(t.east)-(.15,0)$);
\draw[->] ($(t.south)+(0,.15)$) to (o);
\draw[->] (bt) to (at);
\draw[->] (bp) to (ps);

\node at ($(t.center)-(2.5,-.5)$) {\footnotesize $\begin{cases}1,\ a(t)\geq T\\ 0,\ a(t)< T\\\end{cases}$};
\node at ($(t.center)+(2.6,.5)$) {\footnotesize $a(t)e^{j(\phi(t)+\theta)}$};
\end{tikzpicture}
        \caption{A nonlinear unit cell schematic leading to the designed amplitude step activation: The input signal undergoes a combination of amplitude thresholding with a threshold controlled by an external bias $B_T$ and phase shifting controlled by an external bias $B_\theta$.}
        \label{fig:1a}
    \end{subfigure}
    \hfill
    \begin{subfigure}{0.49\textwidth}
        \includegraphics[width=\linewidth]{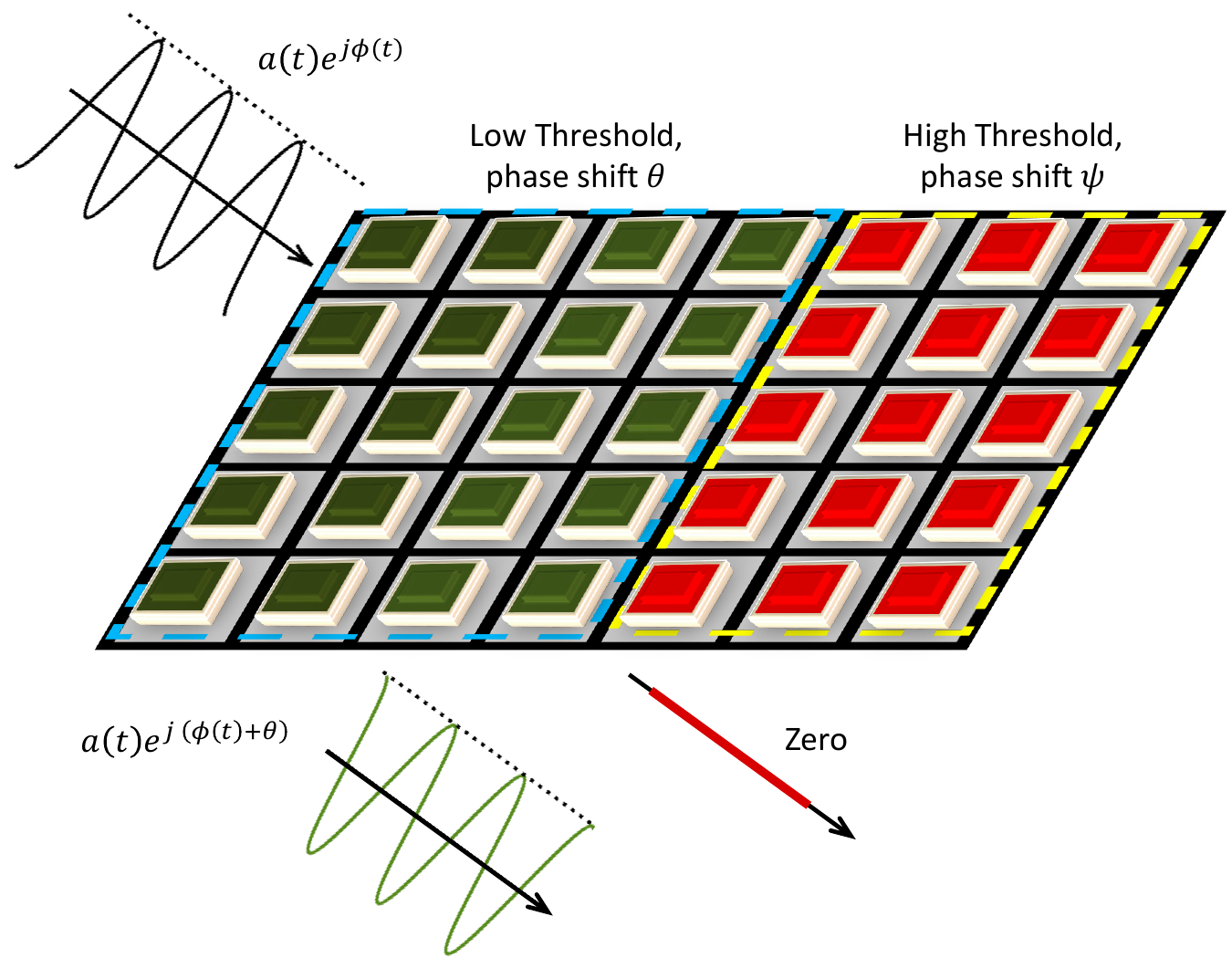}
        \caption{A layer of an NL-SIS: In this example, the layer is configured as follows: a low threshold region (left) with an output signal being the input signal with a phase shift of $\theta$, and a high threshold region (right) with a zero output signal.}
        \label{fig:1b}
    \end{subfigure}
    \caption{Conceptual illustration of an amplitude-dependent NL-SIS.}
    \label{UnitSchematic}
\end{figure*}

Before presenting the unit cell architecture, we set the following design conditions following the proposed function $f(x(t))$ above: (i) the unit cell should allow transmission of EM waves only when the incident amplitude exceeds a defined threshold, (ii) this threshold should be tunable, (iii) the unit cell provides phase control independent from the amplitude-thresholding, and (iv) the architecture must be passive (no amplitude gain). Conceptually, such a unit functions as a threshold-controlled absorber.

The proposed unit cell architecture can comprise one or multiple elements with mounted lumped components, such as varactors and/or PIN diodes. When the incident EM wave amplitude is below the defined threshold, the elements are biased such that the unit behaves as an absorber. Once the input amplitude exceeds the threshold, the unit transitions to a transmissive mode, applying an externally-controllable phase shift. A reflective surface with a conceptually similar operation was demonstrated in \cite{active_non_linear}, where a multi-layer structure containing a varactor-based element and RF signal power detection was shown to switch between different reflection profiles depending on the signal amplitude, which controls the applied bias on the varactor.

Fig.~\ref{UnitSchematic} illustrates the proposed NL-SIS, where the response of the surface depends jointly on the signal amplitude, threshold control bias $B_T$, and phase control bias $B_\theta$, and the nonlinearity is introduced through the amplitude-dependent switching behaviour. Note that, unlike \cite{active_non_linear}, our architecture eliminates the need for detection circuitry, as the threshold and phase are independent parameters, simplifying the overall implementation.

In the next section, we evaluate the system-level performance potential of NL-SIS in wireless communication scenarios.

\section{Advancing Wireless Systems Using NL-SIS}

Despite the promising system-level enhancements enabled by nonlinear surfaces, current implementations remain constrained to linear surfaces. In this section, we explore the capabilities introduced by NL-SIS and subsequently discuss optimization strategies to integrate NL-SIS into wireless communication systems.

\subsection{NL-SIS Capabilities}

As discussed in Sec.~\ref{nl_SIM_theory}, nonlinearity in surfaces can manifest in various forms, including polarization control, amplitude-dependent behaviour, and other nonlinear wave manipulations. In this work, we model the nonlinearity as a step function applied to the amplitude of the input EM wave.

An SIS with an amplitude-step functionality enables the decoupling of amplitude and phase manipulations by the SIS. In wireless systems, the signal's amplitude typically reflects its strength, affected by path loss, fading, and shadowing, while the phase depends on propagation delay, multipath interference, and Doppler shifts. NL-SIS enables independent control and learning of amplitude and phase components, which allows for more optimized operations. This decoupling supports advanced functions such as phase compensation, which becomes more tractable when phase is decoupled from amplitude, and interference mitigation, where phase control enables beam nulling and amplitude control allows beampattern shaping. It also enhances key processes such as channel estimation, equalization, beamforming, and the detection of quadrature amplitude modulated (QAM) signals.

Further, introducing nonlinearity leads to stable gradients and high noise resistance when optimizing the SIS. The nonlinear behaviour in NL-SIS introduces a form of smooth sparsity, where small-magnitude signals are suppressed. This helps stabilize optimization by reducing issues like gradient vanishing or explosion and avoiding saturation or divergence. Additionally, the amplitude-step nonlinearity compresses inputs into bounded ranges, which simplifies detection and estimation schemes. This results in improved robustness to perturbations in the input signal.
In non-Gaussian noise environments, an NL-SIS can emulate the behaviour of nonlinear de-noising techniques such as median or bilateral filters, which often outperform their linear counterparts \cite{filtergus}. Finally, the nonlinearity acts as an implicit regularizer, improving the system's ability to generalize across diverse channel conditions and deployment scenarios.

\subsection{Optimizing an NL-SIS}

Optimizing an SIS remains an active area of research due to the large optimization space and the number of hyperparameters involved. Various approaches have been explored, including gradient-based optimization (requiring differentiable objective functions) and deep reinforcement learning (DRL) approaches \cite{SISComSens}. However, these methods face two major limitations: (i) the final performance is sensitive to the initial parameter settings, often leading to inconsistent results, and (ii) iterative methods are computationally less efficient than backpropagation-based training frameworks in using GPU resources.

To address these challenges, we propose modelling the NL-SIS as a complex-valued neural network (CV-NN) \cite{CVNN}, where each unit cell acts as a neuron equipped with an amplitude step activation function and controllable through a tunable phase shift and amplitude threshold. The optimization objective can be to minimize the SER of end-to-end transmission between the transmitter and receiver, or some other objective, depending on the application. Notably, the underlying deep learning model supports any differentiable objective function, making it adaptable for broader applications such as integrated sensing and communication (ISAC). This modelling offers several key advantages:
\begin{itemize}
    \item Modern deep learning frameworks can automatically compute the gradient of the objective function.
    
    \item GPU-accelerated backpropagation significantly speeds up training algorithms for optimizing the phase shifts and thresholds of the unit cells.
\end{itemize}

The NL-SIS architecture outlined here has the potential to improve wireless system performance across a range of advanced applications, including communication, channel coding, and ISAC (one of which is explored in the following case study).

\section{Case Study}

\begin{figure*}
\centering
\includegraphics[width=\linewidth]{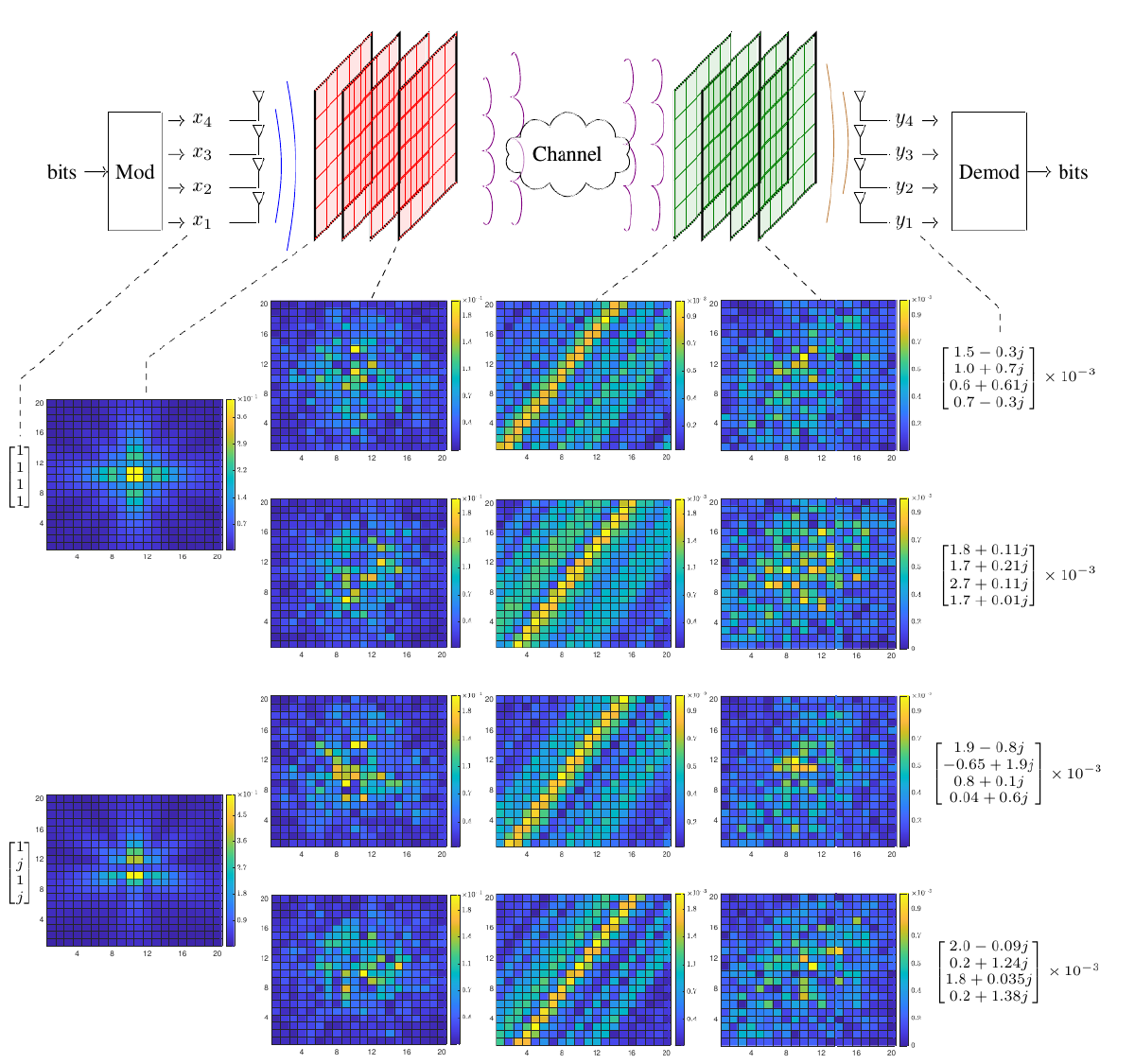}
\caption{An SIS-assisted MIMO system with 4 antennas transmitting two QPSK Signals. The heat maps show the signal amplitudes across an SIS layer for an L-SIS (top) and an NL-SIS (bottom).}
\label{heatmap}
\end{figure*}

Obtaining perfect CSI in wireless communication systems poses practical challenges, particularly in scenarios involving large antenna arrays or reconfigurable surfaces.
Under perfect CSI and linear-SIS (L-SIS), channel-diagonalization techniques can be used, such as using a singular value decomposition (SVD) of the channel. However, when perfect CSI is unavailable or costly to acquire, this method becomes infeasible. Alternatively, the wireless systems can be optimized based on the end-to-end performance without requiring perfect CSI, but rather statistical CSI. In such cases, channel variations can be treated as part of the noise in the system. An NL-SIS can be trained to perform modulation/detection for the underlying channel exhibiting both thermal noise (additive) and channel noise (multiplicative), thus maintaining robust end-to-end performance.
To demonstrate this, we study a communication scenario in which a transmitter aims to communicate with a receiver, assuming only channel statistics are known. We evaluate the performance of the proposed NL-SIS in comparison with a conventional L-SIS, focusing on its robustness to imperfect channel knowledge and generalization across statistical channel conditions.

\subsection{System Description}

We consider a MIMO communication system incorporating SISs at the transmitter (Tx) and the receiver (Rx) as shown in Fig.~\ref{heatmap} (top figure), with optimization performed using only statistical channel knowledge. The SIS parameters at the Tx and the Rx are tuned using \textit{statistical} channel knowledge (no instantaneous CSI is available) to minimize the SER.
The Tx is equipped with a planar antenna array comprising four elements with spacing $\lambda$ where $\lambda$ is the wavelength at a carrier frequency of 28~GHz. An SIS is deployed parallel to the Tx antenna array and oriented perpendicular to the link axis at a distance of $4\lambda$. The transmitter-SIS has a fixed thickness of $2.5\lambda$ and consists of $L = 4$ stacked layers, each comprising $N$ reconfigurable elements. The Rx is placed 50 meters away from the Tx and features a planar antenna array with four elements, accompanied by a receiver-SIS with the same configuration as the transmitter-SIS.
Each unit cell in the transmitter-SIS and receiver-SIS has fixed dimensions of $\lambda/2 \times \lambda/2$. In the L-SIS case, these elements apply a controllable phase shift to the incident EM wave. In the NL-SIS case, each element applies amplitude thresholding and phase shifting.

The Tx transmits a vector $\mathbf{x}$ of 4-QAM symbols from the set $\{1,j,-1,-j\}$ with power $1$~dBm. We model the channel ($\mathbf{H}$) between the last layer of the transmitter-SIS and the first layer of the receiver-SIS as a Rician fading channel, whereas the channels between the Tx antennas and the last transmitter-SIS layer, as well as the first receiver-SIS layer and the Rx antennas, are modelled using Rayleigh-Sommerfeld diffraction theory.
The received signal $\mathbf{y}$ can be written as

\begin{equation}
    \mathbf{y} = \mathbf{P}(\mathbf{HQ}(\mathbf{x})) + \mathbf{n}, 
\end{equation}
where $\mathbf{Q}(\cdot)$ is the transmitter-SIS nonlinear operation which maps from $\mathbb{C}^4$ to $\mathbb{C}^N$, $\mathbf{P}(\cdot)$ is the receiver-SIS nonlinear operation which maps from $\mathbb{C}^N$ to $\mathbb{C}^4$, and $\mathbf{n}$ is a noise vector with independent Gaussian noises with a power spectral density of $-120$~dBm/Hz.

The goal of the SIS is to shape the waveforms at both SIS ends to minimize the SER. To train the SIS parameters, we generate an artificial dataset of noise vectors and channel realizations that follow the known noise and channel statistics (which depend on the known Tx and Rx locations and Rician factor), respectively. Then, we train the SIS using these channels, noises, and random 4-QAM vectors by minimizing the SER. The results are discussed next.

\subsection{Convergence and Generalization}

\begin{figure}
\centering
\tikzset{every picture/.style={scale=.8}, every node/.style={scale=0.9}}
\input{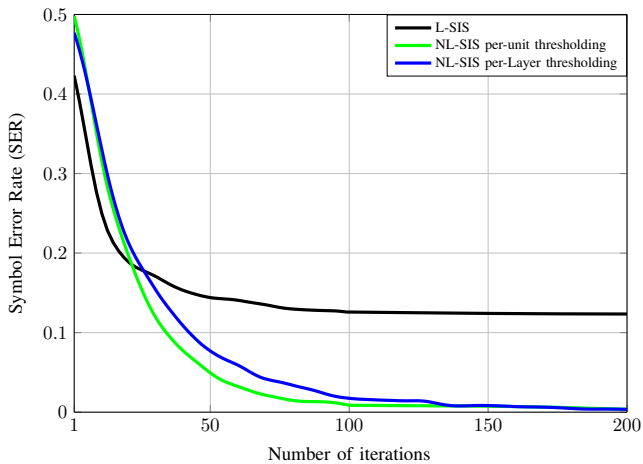}
\caption{Symbol error rate versus the number of iterations for L-SIS, NL-SIS with per-unit threshold optimization, and NL-SIS with per-layer threshold optimization.}
\label{elements_res}
\end{figure}

We start by examining the convergence of the training of the SIS-based NN model and the performance improvement achieved with NL-SIS. For this analysis, we fix the number of elements per layer to $N=400$ and use a Rician factor $\kappa = 10$. After training, both L-SIS and NL-SIS are tested on previously unseen data to assess generalization.

When training, we consider per-unit thresholds where each unit cell has its threshold to be optimized, and per-layer thresholds where all unit cells in a given layer have the same threshold. Fig.~\ref{elements_res} shows the SER as a function of training iterations for L-SIS and NL-SIS. 
When the number of iterations is low, the L-SIS has the steepest training slope due to its lower number of trainable parameters $N \times L \times 2$, where we multiply by 2 as we have SIS at both the Tx and Rx side. The NL-SIS with per-unit thresholding has approximately twice as many parameters. Interestingly, over iterations, per-layer thresholding has a lower training slope compared with the per-unit thresholding despite having fewer parameters, specifically $(N + 1) \times L \times 2$, because tuning a single threshold per layer requires finding a common value that works well across all units in that layer.
In terms of performance, NL-SIS significantly outperforms L-SIS with an SER of 0.1236 for the L-SIS and of 0.0035 for the NL-SIS. Moreover, per-layer thresholding results in an SER only 2.8\% higher than per-unit thresholding, suggesting that per-layer control may offer a good trade-off between complexity and performance. This insight is valuable from a hardware perspective, as implementing per-layer thresholding simplifies the biasing circuitry compared to per-unit designs.

To understand the reason why the NL-SIS outperforms the L-SIS, let us consider the heat maps plotted in Fig.~\ref{heatmap}. Here, we present maps of the signal amplitude at the input and output of the L-SIS and NL-SIS for various transmitted signals. Each sub-figure displays the amplitude distributions across four layers with the input and output of the transmitter-SIS (first and last layers), followed by the input and output of the receiver-SIS (also first and last layers), all arranged from left to right.

The heat maps show that the NL-SIS creates signal patterns at the transmitter-SIS output that lead to a stronger and more concentrated signal at the input of the receiver-SIS compared to the L-SIS case. They also show that the NL-SIS at the receiver processes the received signal into a signal with more high-amplitude values (showing as bright spots in the heat map) compared to the L-SIS. This leads to stronger received QAM symbols, resulting in higher resilience to noise and, consequently, lower SER. We note that this behaviour was observed consistently in multiple instances of transmitted/received signals with random channel realizations, which are not shown here due to space limitations. 

The superiority of the NL-SIS here could stem from its ability to switch off some of its unit-cells (via threshold configuration). This feature enables the NL-SIS to null some unit-cells' signals which would otherwise negatively impact the signal at the next SIS layer; an operation which is not possible in the L-SIS case.

\subsection{Noise Resistivity}

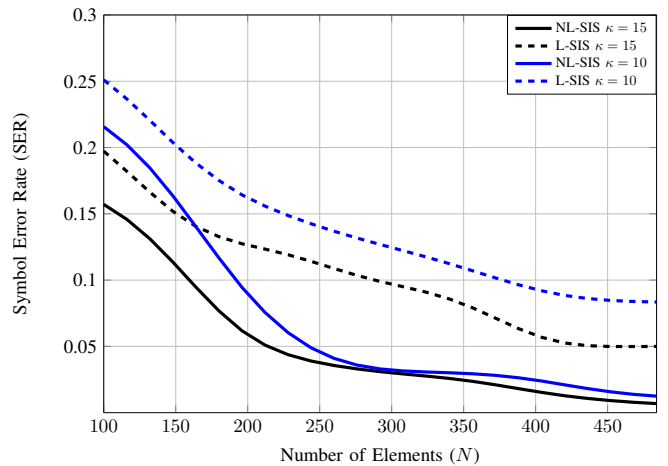
\begin{figure}
\centering
\tikzset{every picture/.style={scale=.8}, every node/.style={scale=0.9}}
%
%

\definecolor{mycolor1}{rgb}{0.00000,0.44700,0.74100}%
\definecolor{mycolor2}{rgb}{0.85000,0.32500,0.09800}%
\begin{tikzpicture}

\begin{axis}[%
grid=major,
width=4.521in,
height=3.255in,
scale only axis,
xmin=100,
xmax=484,
xlabel style={at={(axis cs:292,0)},anchor=north},
xlabel={Number of Elements ($N$)},
separate axis lines,
every outer y axis line/.append style={black},
ymin=0,
ymax= 0.3,
ytick={0.05,0.1,0.15,0.2,0.25,0.3},
yticklabels={0.05,0.1,0.15,0.2,0.25,0.3},
ylabel style={font=\color{black},at={(axis cs:100,0.15)},anchor=south},
ylabel={Symbol Error Rate (SER)},
axis background/.style={fill=white},
xmajorgrids,
ymajorgrids,
legend style={{at =  (axis cs:484,0.3)}, nodes={scale=0.75, transform shape}, anchor=north east, legend cell align=left, align=left, draw=black}
]

\addplot [color=black, line width=1.5pt]
  table[row sep=crcr]{%
100	0.15712900340984\\
116	0.145840441231677\\
132	0.131212917438938\\
148	0.113640888986233\\
164	0.0946962367331452\\
180	0.0768030601139038\\
196	0.0617855123976815\\
212	0.0509818748428178\\
228	0.0437516707821359\\
244	0.0389584271442778\\
260	0.0355857450572459\\
276	0.0330074201804419\\
292	0.0309351840264965\\
308	0.0292085377470934\\
324	0.0276024172226746\\
340	0.0258307147526993\\
356	0.0236566631560704\\
372	0.0210328576164493\\
388	0.0181523895338309\\
404	0.0153283876673284\\
420	0.0127738821188719\\
436	0.0107436245614098\\
452	0.00911253353455512\\
468	0.00782859473033573\\
484	0.00684353702558138\\
};
\addlegendentry{NL-SIS $\kappa = 15$}
 
\addplot [color=black,dashed, line width=1.5pt]
  table[row sep=crcr]{%
100	0.197247391101401\\
116	0.182106161571183\\
132	0.166387531478921\\
148	0.151919660239025\\
164	0.140522515743482\\
180	0.132774316704849\\
196	0.127409551254799\\
212	0.123399363733225\\
228	0.119139982056765\\
244	0.114152558303137\\
260	0.108736811059054\\
276	0.103517261274928\\
292	0.0989594471261843\\
308	0.0949755366879602\\
324	0.0908826945141096\\
340	0.085783765858716\\
356	0.0791628256412362\\
372	0.0713293463392224\\
388	0.0634351841119518\\
404	0.0569047898275146\\
420	0.0525750222211121\\
436	0.0505177499666264\\
452	0.0499020063830478\\
468	0.0498849331572482\\
484	0.0499499313538913\\
};
\addlegendentry{L-SIS $\kappa = 15$}

\addplot [color=blue, line width=1.5pt]
  table[row sep=crcr]{%
100	0.215728057354862\\
116	0.202150006852156\\
132	0.184694210950266\\
148	0.163618157574897\\
164	0.140285916508798\\
180	0.11674143140036\\
196	0.0945484838678301\\
212	0.0756229282533075\\
228	0.0603928744312613\\
244	0.0489282952474757\\
260	0.0409505207362972\\
276	0.0358861836659818\\
292	0.0329840218101004\\
308	0.0314709918201454\\
324	0.0306810698775789\\
340	0.0301006580158469\\
356	0.0293485515935274\\
372	0.0281506476933404\\
388	0.0263423987518995\\
404	0.023904704734134\\
420	0.0210082077703014\\
436	0.0182024463963715\\
452	0.0157337765283553\\
468	0.0138149629072732\\
484	0.0124558473373588\\
};
\addlegendentry{NL-SIS $\kappa = 10$}
 
\addplot [color=blue,dashed, line width=1.5pt]
  table[row sep=crcr]{%
100	0.251029502443126\\
116	0.236684482614428\\
132	0.220535305164777\\
148	0.203841079704187\\
164	0.188311138844755\\
180	0.17517190625386\\
196	0.164316368629451\\
212	0.155818077378802\\
228	0.148718818085148\\
244	0.142483913152159\\
260	0.136855491634503\\
276	0.131695263121041\\
292	0.126868944168221\\
308	0.122188586963349\\
324	0.117429673802247\\
340	0.11241271247783\\
356	0.10710367434661\\
372	0.101676419507557\\
388	0.0964954489545583\\
404	0.0919882581749068\\
420	0.088465501825844\\
436	0.0861004254311475\\
452	0.0846586126167121\\
468	0.0838642883492095\\
484	0.0834556266154445\\
};
\addlegendentry{L-SIS $\kappa = 10$}

\end{axis}

\end{tikzpicture}%
\caption{Symbol error rate versus the number of SIS elements for L-SIS and NL-SIS under different values of $\kappa$.}
\label{elementsKappa}
\end{figure}

As previously shown in the literature, increasing the number of unit cells in an SIS enhances its beamforming gain, which improves system performance in terms of achievable rate and channel estimation accuracy, for instance \cite{gradientdecentSIMComm}. In this subsection, we evaluate the effect of the number of unit cells in an L-SIS and NL-SIS on their adaptability and robustness to noise and channel imperfections for different values of $\kappa$.
Fig.~\ref{elementsKappa} shows the SER versus the number of SIS elements $N$ for Rician factors $\kappa = 10$ and $\kappa = 15$. As expected, increasing $N$ improves the SER for both architectures. Specifically, for $\kappa = 15$, the SER decreases from 0.184 to 0.004 as $N$ increases from 100 to 484 in the NL-SIS case, and decreases from 0.254 to 0.048 in the L-SIS case. This improvement highlights how larger SISs can be reconfigured better for statistical channel characteristics.
Notably, L-SIS performance saturates beyond $N = 400$, indicating that further increasing the elements yields limited gains. In contrast, NL-SIS continues to benefit from added elements, thanks to its increased expressive power due to the nonlinear units. This scalability allows NL-SIS to better adapt to imperfect channel information. Moreover, as $\kappa$ decreases from 15 to 10, indicating a more challenging channel environment, the NL-SIS maintains a significant performance edge over L-SIS, particularly when $N > 300$. This illustrates the superior adaptability of NL-SIS to varying channel conditions.

The example above shows the benefits of using an NL-SIS in a communication system, while maintaining manageable design complexity and power consumption. Next, we discuss some challenges and opportunities for future research.

\section{Opportunities and Challenges}

The integration of NL-SIS into wireless systems opens up promising avenues for performance enhancement across different applications, while also introducing new technical challenges. This section highlights key application scenarios where an NL-SIS can offer significant benefits, as well as the practical challenges that may arise in its deployment.

\subsection{NL-SIS-Assisted low resolution MIMO}

Power consumption in massive MIMO systems is primarily related to the active components connected to the antennas, including power amplifiers, analog-to-digital (ADC) converters, and digital-to-analog (DAC) converters. Typically, implementing a low-resolution ADC and DAC reduces the power consumption, but decreases the system performance \cite{onebitmimo}. NL-SIS presents a promising solution by enabling nonlinear signal processing that can enhance the effective resolution of low-resolution MIMO systems. One potential implementation of NL-SIS is enhancing the rate of low-resolution MIMO systems by using its nonlinear operations to produce and process codes with higher resolution. For example, integrating an NL-SIS with a one-bit MIMO system, which only allows two discrete states for the real and imaginary parts per antenna, can result in a transmitted signal with richer states, thus increasing reliability (lowering error rates).

Moreover, an NL-SIS can improve power efficiency in MIMO systems by acting as a nonlinear encoder or decoder, where antennas serve as energy detectors and amplitude modulators. In other words, each antenna may encode/detect 1 bit using two amplitude levels, with the NL-SIS converting between amplitude levels and complex-valued signals. This reduces the hardware complexity and power consumption while preserving performance.

\subsection{NL-SIS-Assisted Sensing}

Integrating SIS into wireless systems offers significant potential for improving sensing functions such as target detection and localization, due to SIS ability to perform multi-stage wavefront processing. However, conventional L-SIS architectures typically emulate antenna arrays with uniform amplitude excitation. The resulting sensing beams, while capable of producing narrow half-power beamwidths, suffer from high sidelobe levels, which degrade accuracy under noisy conditions.
Additionally, traditional sensing systems often assume perfect channel knowledge during beam generation. Therefore, even minor phase estimation errors can cause substantial signal degradation, compromising detection performance. NL-SIS offers a compelling solution to these limitations. By enabling non-uniform amplitude transformations, an NL-SIS can generate beams with reduced sidelobes—thereby improving localization and detection accuracy.
As demonstrated in Fig.~\ref{elementsKappa}, an NL-SIS also exhibits superior adaptability to channel imperfections compared to L-SIS, making it particularly advantageous in realistic, noisy, and dynamic environments.

\subsection{Beyond Nonlinear NN}

While this study focuses on modelling SIS nonlinearity using NN principles via amplitude thresholding at the unit cell or layer level, other forms of nonlinear operations are also feasible within the SIS framework, as discussed in Sec.~\ref{Sec.Nonlinear}.
For example, polarization conversion can be employed in wireless systems to simplify antenna design by transforming linear polarization to other types, or to enable polarization filtering to mitigate interference and polarization contamination. Moreover, nonlinear phase or amplitude responses can be designed as functions of frequency rather than power. This frequency-dependent behaviour could help reduce beam squinting, thereby improving performance in wideband systems.

\subsection{Advancing Secure Communication}

Reconfigurable surfaces such as RIS have been extensively explored for physical-layer security, enabling secure key generation and enhancing link confidentiality by either suppressing signal reception at eavesdroppers or increasing artificial noise while boosting the signal at legitimate users. 
Although SISs are designed with a different purpose from RIS, in the context of physical-layer security, NL-SIS offers a promising solution for establishing secure links without relying on complex algorithms.
An NL-SIS can implement nonlinear mappings at the receiver, which are inherently more difficult to predict and reverse-engineer than linear transformations. Moreover, compared to L-SIS, NL-SIS structures can generate radiation patterns with lower sidelobe levels and configurable nulls, thereby reducing signal leakage to unintended receivers.

\section{Conclusion}\label{conclusion}

Nonlinearity is a fundamental phenomenon in RF circuit design and electromagnetic surface engineering. In EM surfaces, nonlinearity can be achieved by integrating nonlinear components and operating them in their nonlinear regime or by applying nonlinear control mechanisms to the surface. Within the context of SIS, nonlinearity can be introduced through various approaches to optimize the surface response for specific applications.
Motivated by the analogy between SIS and NNs, we proposed a unit cell schematic capable of exhibiting step-like nonlinear behaviour. While this study focused on amplitude-thresholding responses, other nonlinear operations can also be implemented by adapting the unit cell architecture to the desired function. As the presented case study shows, nonlinearity in an SIS can decrease the detection SER. We expect similar gains in other wireless communications objectives and applications, but this remains to be seen in future research in this direction.

\bibliographystyle{IEEEtran}
\bibliography{IEEEabrv,bibliography}

\end{document}